\documentclass[a4paper]{article}

\usepackage{INTERSPEECH2020}
\usepackage{multirow}
\usepackage{booktabs}
\usepackage{mathrsfs}
\usepackage{amsmath}
\usepackage{pifont}
\usepackage[marginal]{footmisc}
\usepackage[normalem]{ulem}
\useunder{\uline}{\ul}{}

\title{The NTNU System at the Interspeech 2020 Non-Native Children’s Speech\\ ASR Challenge}
\name{Tien-Hong Lo, Fu-An Chao, Shi-Yan Weng, Berlin Chen}
\address{
  National Taiwan Normal University, Taiwan}
\email{\{teinhonglo, 60747002S, 40547041S, berlin\}.ntnu.edu.tw}

\begin{document}

\maketitle

\begin{abstract}
  This paper describes the NTNU ASR system participating in the Interspeech 2020 Non-Native Children’s Speech ASR Challenge supported by the SIG-CHILD group of ISCA. This ASR shared task is made much more challenging due to the coexisting diversity of non-native and children speaking characteristics. In the setting of closed-track evaluation, all participants were restricted to develop their systems merely based on the speech and text corpora provided by the organizer. To work around this under-resourced issue, we built our ASR system on top of CNN-TDNNF-based acoustic models, meanwhile harnessing the synergistic power of various data augmentation strategies, including both utterance- and word-level speed perturbation and spectrogram augmentation, alongside a simple yet effective data-cleansing approach. All variants of our ASR system employed an RNN-based language model to rescore the first-pass recognition hypotheses, which was trained solely on the text dataset released by the organizer. Our system with the best configuration came out in second place, resulting in a word error rate (WER) of 17.59 \%, while those of the top-performing, second runner-up and official baseline systems are 15.67\%, 18.71\%, 35.09\%, respectively. 
\end{abstract}

\section{Introduction}

Due to the rapid advancements in automatic speech recognition (ASR) with various sophisticated deep neural network (DNN) modeling techniques, alongside the availability of large amounts of training data and powerful computational resources, there has been widespread adoption of ASR solutions in many application domains such as personal assistants, interactive voice responses (IVR) and among others, with which people can interact naturally with machines using their voice.

Although some current top-of-the-line ASR systems can even reach the performance level of professional human annotators in specific conditions [1, 2], many real-world application scenarios still pose great challenges for ASR. One of the most challenging application scenarios is recognition of non-native children's speech, for which two sets of intricate phenomena coexist, often dramatically reducing ASR performance. One is the non-native pronunciation behaviors, including mispronounced words, ungrammatical utterances, code-switched words, and disfluencies. The other is the linguistic differences of children from adult speech at many levels, including acoustic, prosodic, lexical, morphosyntactic, and pragmatic levels, to name a few [3]. This may also manifest in the inter- and intra-speaker variability of children due to varying vocal tract lengths and undeveloped pronunciation skills [4-7]. What is more, the scarcity of publicly available large-scale non-native children’s speech data with human annotations further hamper the ASR performance.

\begin{figure}[t]
  \centering
  \includegraphics[width=\linewidth]{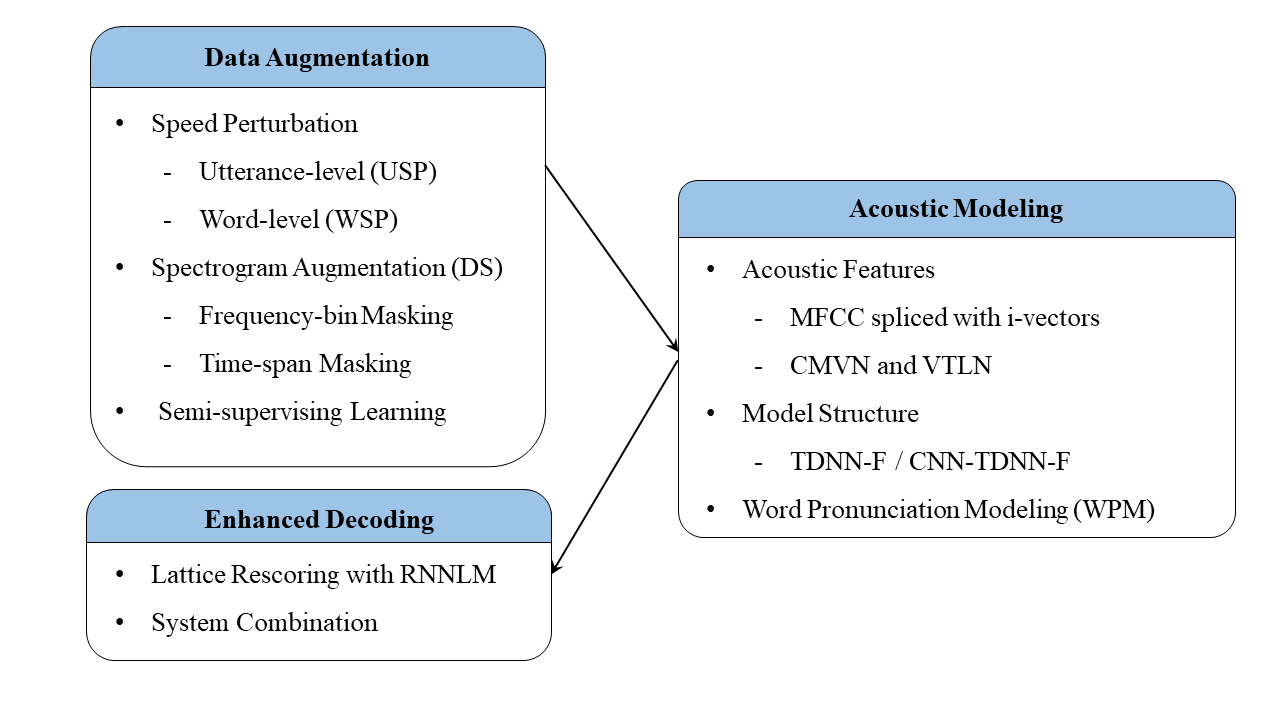}
  \caption{Highlight of the NTNU ASR system configuration.}
  \label{fig:workflow}
\end{figure}

This paper describes the NTNU ASR system participating in the Interspeech 2020 Non-Native Children’s Speech ASR Challenge (TLT-school Challenge) supported by the SIG-CHILD group of ISCA\footnote{\noindent https://sites.google.com/view/wocci/home/interspeech-2020-special-session}. Due to the coexisting diversity of non-native and children speaking characteristics, this ASR shared task is made much more challenging. In the setting of closed-track competition, all participants were restricted to develop their systems solely based on the training speech and text corpora provided by the organizer. To deal with this under-resourced issue, we built our ASR system on the basis of a top-of-the-line, hybrid deep neural network and hidden Markov model (DNN-HMM) structure for acoustic modeling, with the lattice-free maximum mutual information (LF-MMI) criterion [8] for model optimization. More specifically, the DNN architecture involves several layers of convolutional neural network (CNN) followed by several layers of factorized time-delay neural network (TDNNF) [9], holistically denoted by CNN-TDNNF hereafter. In order to combat the data-sparsity and high variability of non-native children’s speech for robust acoustic modeling, we augmented the given training dataset with several spectrogram- and speed perturbation-based data augmentation strategies, including the recently proposed spectrogram augmentation (denoted by SpecAugment) method [10], and both utterance- and word-level speed perturbations [11] in the training phase. Furthermore, inspired by [5], speech feature extraction was conducted with the aid of vocal tract length normalization (VTLN) [12], as well as cepstral mean and variance normalization (CMVN) [13]. Apart from the above, we capitalized on the so-called word pronunciation modeling [14] in place of the conventional pronunciation modeling approach [5]. All variants of our ASR system employed a recurrent neural network (RNN)-based language model (denoted by RNNLM) to rescore the first-pass recognition hypotheses [15], in conjunction with a lattice combination procedure [16]; the RNNLM model was trained solely on the text dataset provided by the organizer. The synergy of all abovementioned treatments brought about a significant improvement over the baseline system announced by the organizer. Figure 1 outlines the configuration of our system.

The remainder of this paper is organized as follows: Section 2 sheds light on the strategies that were employed for training data cleansing and augmentation. Section 3 presents the details of the acoustic modeling process. Section 4 describes the RNN-based language model as well as the accompanying lattice rescoring methods. After that, the experimental setup, results and discussion are given in Section 5. We conclude the paper and envisage future research directions in Section 6.

\section{Data Cleansing and Augmentation}

\subsection{Data Cleansing}

Hybrid DNN-HMM (e.g., CNN-TDNNF) acoustic models have shown to be significantly superior than the conventional HMM-based acoustic models that employ Gaussian mixture models (GMM) to characterize the emission probabilities of frame-level speech feature vectors being generated by each HMM state (denoted by GMM-HMM) on many ASR tasks. Hybrid DNN-HMM acoustic models still have to resort to GMM-HMM acoustic models to obtain good forced-alignment information for better estimating their corresponding neural network parameters. Therefore, the GMM-HMM acoustic model of our best system was training with the audio segments selected out from the speech training dataset with high recognition confidence scores generated by an existing hybrid DNN-HMM system. As we shall see later, the empirical ASR results confirm this intuitive data-cleansing therapy.

Due to the constraint posed by the closed-track competition, viz. only the speech and text corpora provided by the organizer could be used for the ASR system development, we thus set out to leverage different data-augmentation strategies based on label-preserving transformations, including both utterance- and word-level speed perturbation and spectrogram augmentation, to diversify and enrich the original speech training dataset, apart from the aforementioned data-cleansing operation. We anticipated that these data-augmentation strategies could further push the performance limit of our ASR system.

\subsection{Utterance- and Word-level Speed Perturbation}
To alleviate the data-scarcity problem for acoustic modeling, a natural thought is to perform utterance-level speed perturbation [11]. It modifies the speaking rate of a speech utterance by resampling its waveform signal. Following the procedure described in [11], in this paper two additional copies of the original speech training data were created by perturbing the speaking rate of each training utterance to 0.9 times and 1.1 times of its original one, respectively. In this way, the training data had increased three-fold. 

Furthermore, in initial experiments, we observed that the word-level speech of non-native children’s utterances exhibits high inter- and intra-speaker variabilities and thus tends to be unstable. To capitalize on this observation, we proposed a word-level speed perturbation method so as to make the resulting acoustic models better accommodate the intricate pronunciation phenomena inherent in non-native children’s speech. Word-level speech perturbation was conducted in a two-stage manner. At the first stage, word-level boundaries of the original training utterances were obtained with a baseline hybrid DNN-HMM ASR system. At the second stage, the speaking rate of each word segment was perturbed by randomly altering it to 0.9 times or 1.1 times of the original one. More specifically, one copy of the training dataset had 80\% of its word segments increase their speaking rate to 1.1 times and 20\% of its word segments reduce their speaking rate to 0.9 times of the original ones. Alternatively, another copy of the training dataset had 20\% of its word segments increase their speaking rate to 1.1 times and 80\% of its word segments reduce their speaking rate to 0.9 times of the original ones. To recap, the aforementioned utterance- and word-level speed perturbation procedures will generate four additional copies of training data, as schematically depicted in Figure 2. Note also here that, due to these augmentation operations will change in the lengths of the wave signals, the forced-alignment information of the speed-perturbed utterances were generated using the baseline hybrid DNN-HMM system.

\begin{figure}[t]
  \centering
  \includegraphics[width=\linewidth]{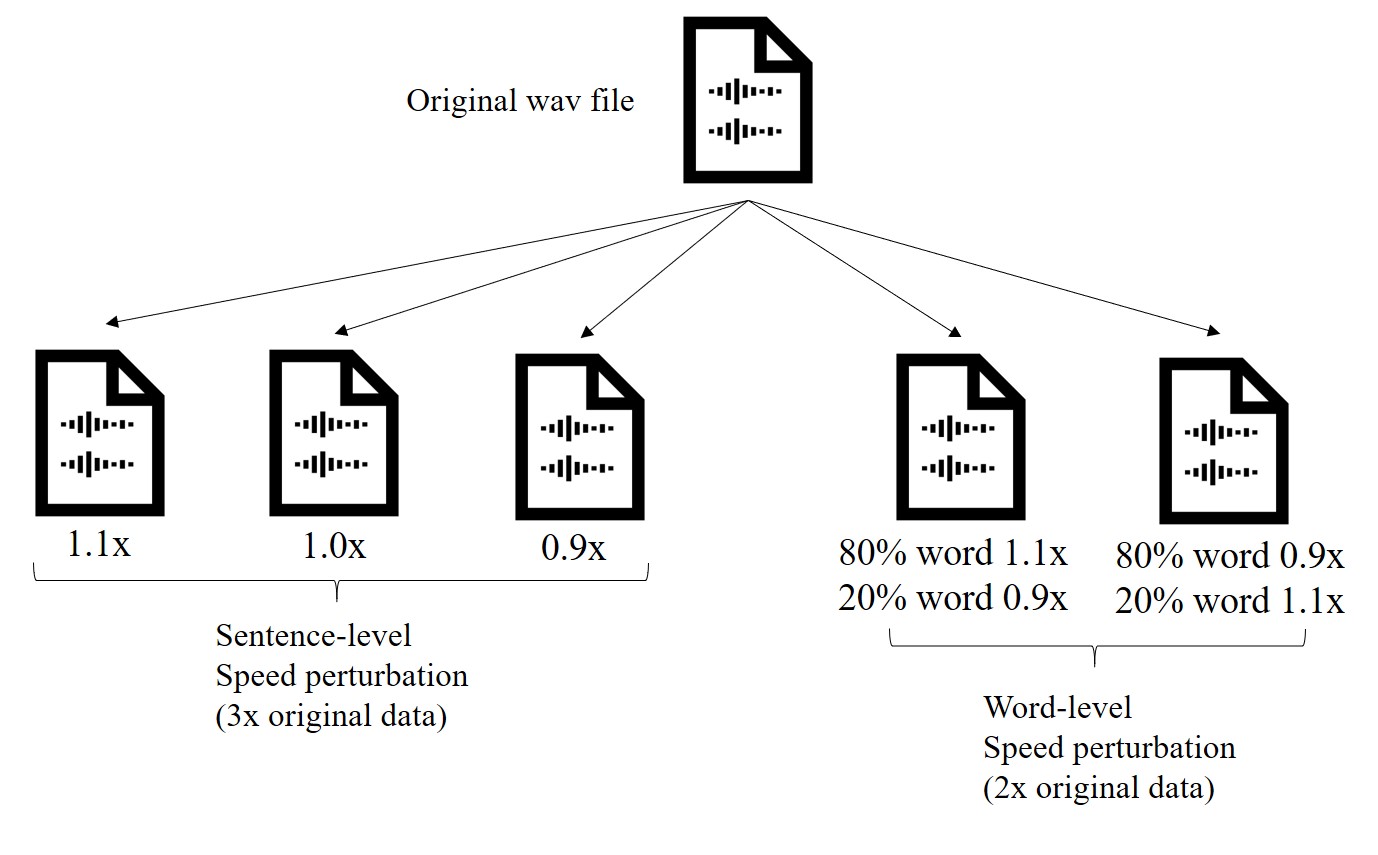}
  \caption{A schematic depiction of the utterance- and word-level speed perturbation procedures.}
  \label{fig:perturbation}
\end{figure}

\subsection{Spectrogram Augmentation}

Another line of research on training data augmentation for ASR acoustic modeling has focused on feature-space augmentation, which takes inspiration from the success of augmentation methods employed in the computer vision (CV) community, many of which augmented CV datasets by adding transformed sample instances along with their respective original labels [17-19]. The most celebrated feature-space augmentation method adopted for acoustic modeling is vocal tract length perturbation (VTLP) [20]. VTLP, which employs a linear warping transformation along the frequency bins, simulates the effect of altering the vocal tract lengths of speakers that produce the training utterances. Very recently, SpecAugment has drawn much attention from the ASR community, which treats the spectrogram of an utterance as an image, and in turn warps it along the time axis, mask blocks of consecutive frequency along the time axis bins and mask the whole frequency bins in short spans of time [10]. These operations collectively lead to considerable word error rate reductions on several benchmark tasks. Apart from the waveform-domain speed perturbation (viz. utterance- and word-level speed perturbation) mentioned previously in Section 2.2, SpecAugment was also applied to generate augmented acoustic training data. To this end, we made use of the component ‘spec-augment-layer’ of Kaldi toolkit [21], which consists only of the operations that mask blocks of consecutive frequency along the time axis bins and mask the whole frequency bins in short spans of time. This is probably because warping spectrogram along the time axis is conceptually similar to waveform-domain speed perturbation, but its costs a great amount of computation and does not get any significant improvement [10].

\section{Acoustic Modeling}
Mel-frequency cepstral coefficients (MFCC) of 40 dimensions, spliced with i-vectors of 100 dimensions [22] were adopted as the frame-level acoustic feature vectors to be fed to the ASR system. VTLN and the cepstral mean and variance normalization (CMVN) operation were conducted in tandem during the feature extraction process. We also observed in our initial experiments that performing VTLN merely on the test dataset yielded better word error rate (WER) results than performing VTLN on the training and test datasets jointly.

As to acoustic modeling, the DNN architecture involves several layers of TDNNF stacking on top of several layers of CNN [9] (cf. Section 1). TDNNF is a subsequent extension of TDNN (time-delay neural network), with the purpose of obtaining better modeling performance and meanwhile reducing the number of parameters by factorizing the weight matrix of each TDNN layer into the corresponding product of two low-rank matrices [9]. It is argued that we can still retain salient information when projecting a weight matrix from a high-dimensional space to low-dimensional spaces by adding a semi-orthogonal constraint to the first low-rank matrix. As an aside, we also incorporated skip connections [23] into TDNNF so as to deepen the network while alleviating the vanishing gradient problem. 

The objective function for training the acoustic model is lattice-free maximum mutual information (LF-MMI) [8]: 
\begin{equation}
    \mathcal{F}_{\mathrm{LFMMI}}=\sum^J_{j=1} \mathrm{log} \frac{P(\mathbf{O}_j|L_j)^k \sum_i P(L_j)}{P(\mathbf{O}_j|L_i)^k P(L_i)}
\end{equation}
Where $\mathbf{O}_j$ and $L_j$ are the acoustic feature vector sequence and the corresponding phone sequence of the $j$-th training utterance, $k$ is a weighting factor, and $P(L_j)$ is the phone $n$-gram language model probability. On the other hand, we use the word-level pronunciation modeling method proposed in [14], in substitution to that the conventional approach proposed in [5]. The former has been proved effective to distinguish multiple word pronunciations and avoid increasing the confusability of the vocabulary. Among other things, we observed experimentally that modeling the probability of inserting silence boundaries for word-level pronunciations in an explicit manner could bring about additional performance gains. 

\section{Language Modeling}
A recurrent neural network language model (RNNLM) instantiated with a forward long short-term memory (LSTM) [15] architecture was trained on the text dataset provided by organizer. The local training objective of RNNLM at word position l in the text dataset is expressed by: 
\begin{equation}
    \mathcal{F}_{\mathrm{RNNLM}}=z_l+1-\sum_i \mathrm{exp} z_i
\end{equation}
where $z_l$ denotes the logit of RNNLM at word position $l$. According to [15], this objective function can be viewed as an approximation of the conventional cross-entropy objective function, which, however, can speed up the training process (viz. the inference time) by allowing for a sampling method to accelerate the training convergence. RNNLM was used for the second-pass lattice-rescoring [15], in conjunction with a word $n$-gram language model previously used in the first-pass decoding. This word $n$-gram language model was also trained solely on the text dataset provided by the organizer.

\begin{table}[]
\centering
\caption{Statistical information of TLT-school corpus.}
\begin{tabular}{cccc}
\hline
              & Hours & \#Utterances & \#Speakers \\ \hline
Train (full)  & 49    & 13,999       & 340        \\ \hline
Train (small) & 32    & 7,370        & 340        \\ \hline
Development   & 2     & 562          & 84         \\ \hline
Evaluation    & 2     & 578          & 84         \\ \hline
\end{tabular}
\end{table}

\begin{table}[]
\centering
\caption{WER (\%) results on the development dataset with the baseline acoustic models trained on the small-sized training dataset; * indicates that the model was trained on the full training dataset instead.}
\begin{tabular}{cccc}
\hline
\multirow{2}{*}{Acoustic Model} & \multirow{2}{*}{DC} & \multirow{2}{*}{WPM} & WER (\%)        \\ \cline{4-4} 
                                &                     &                      & Development Set \\ \hline
TDNNF                           & -                   & -                    & 26.41           \\ \hline
TDNNF                           & \ding{51}           & -                    & 23.13           \\ \hline
CNN-TDNNF                       & \ding{51}           & -                    & 22.34           \\ \hline
CNN-TDNNF                       & \ding{51}           & \ding{51}            & 21.75           \\ \hline
CNN-TDNNF*                      & \ding{51}           & \ding{51}            & 21.20           \\ \hline
\end{tabular}
\end{table}

\section{Experiments}
\subsection{Experimental Setup}
We evaluated our approaches to low-resourced non-native children’s English speech ASR on the TLT-school corpus [24]
, while the baseline ASR systems was developed with the Kaldi toolkit [25] and the recipes released by organizer. The TLT-school corpus consists of English spoken responses collected from Italian school students between the ages of 9 to 16. Several intricate phenomena of non-native children’s speech, such as mispronounced, code-switched words and linguistic differences between children and adult speech, make this task much more challenging than before. The training set and development set consisted of 13,999 utterances from 340 speakers, and 562 utterances from 84 speakers, respectively. In addition, the evaluation set was composed of 578 utterances from another set of 84 speakers. A smaller-sized training set, which was used for quick tuning of the baseline settings. Table 1 shows some basic statistics of the TLT-school corpus.

\subsection{Data Cleansing and Pronunciation Modeling}
Our first set of experiments on the development set is designed to analytically investigate the effectiveness of data cleansing (DC) and word-level pronunciation modeling (denoted by WPM), previously proposed in Sections 3 and 4, respectively. To this end, two disparate acoustic models, viz. TDNNF and CNN-TDNNF trained with the small-sized training dataset, are respectively employed as the default acoustic model. Three noteworthy points can be drawn from Table 2. First, the application of DC leads to a relative WER reduction of 12.4\% (\textit{cf.} Rows 1 and 2) as TDNNF is used as the acoustic model. Second, when DC is applied, CNN-TDNNF (stacking CNN with TDNNF) can further yield a relative WER reduction of 3.4\% over that using TDNNF in isolation. Third, working in conjunction with WPM, the performance of CNN-TDNNF the based ASR system can be steadily improved, while using the full training dataset (\textit{cf.} the last row of Table 2) instead of the small-sized training dataset further advances the performance. From now on, unless otherwise stated, we will adopt the model configuration determined in the last row of Table 2 for the following experiments. 

\begin{table}[]
\centering
\caption{WER (\%) results on the development dataset with disparate data-augmentation settings.}
\begin{tabular}{ccc}
\hline
\multirow{2}{*}{\begin{tabular}[c]{@{}c@{}}Spectrogram  \\ Augmentation\end{tabular}} & \multirow{2}{*}{\begin{tabular}[c]{@{}c@{}}Speed \\ Perturbation\end{tabular}} & WER (\%)                                                                              \\ \cline{3-3} 
                                            &                                     & Development Set                                                                       \\ \hline
\ding{51}                                   & USP                                 & 19.92                                                                                 \\ \hline
\ding{51}                                   & WSP                                 & 20.57                                                                                 \\ \hline
\ding{51}                                   & USP+WSP                             & 19.80                                                                                 \\ \hline
\ding{51}                                   & USP+WSP                             & \begin{tabular}[c]{@{}c@{}}\textbf{18.86} \\  (Lattice \\ Rescoring)\end{tabular} \\ \hline
\end{tabular}
\end{table}

\begin{table}[]
\centering
\caption{WER (\%) results of our final system on the development and evaluation datasets, with normal supervised learning or semi-supervised learning.}
\begin{tabular}{ccc}
\hline
\multirow{2}{*}{\begin{tabular}[c]{@{}c@{}}Semi-\\ supervised\\ Learning\end{tabular}} & \multicolumn{2}{c}{WER (\%)}                                                                                            \\ \cline{2-3} 
                                                                                       & \begin{tabular}[c]{@{}c@{}}Development\\ Set\end{tabular} & \begin{tabular}[c]{@{}c@{}}Evaluation\\    Set\end{tabular} \\ \hline
-                                                                                      & \textbf{16.70}                                            & 17.79                                                       \\ \hline
\ding{51}                                                                              & 16.74                                                     & \textbf{17.59}                                              \\ \hline
\end{tabular}
\end{table}

\subsection{Data Augmentation}
In the second set of experiments, we turn to assess the impacts of different combinations of data augmentation methods, viz. spectrogram augmentation and speed perturbation (\textit{cf.} Section 2), on the TLT-school task (viz. non-native children’s English speech ASR). Note here that, for speed perturbation, either utterance-level speed perturbation (denoted by USP) or word-level speed perturbation (denoted by WSP), or their synergy are used to expand the training dataset for acoustic modeling. The corresponding results on the development are shown in Table 3. As compared to the last row of Table 2, we can find that all different combinations of spectrogram augmentation and speed perturbation (\textit{cf.} the first three rows of Table 3) can considerably boost the ASR performance, leading to a relative WER reduction of 6.6\% when with the best combination setting. This results also confirm the merits of conducting data augmentation for resource-scarce ASR tasks, such as the TLT-school task studied in this paper. As a side note, if an additional second-pass lattice rescoring is further applied (with a proper combination of RNNLM and the word n-gram language model), the WER of our system on the development set can be further decreased to 18.86\%. 

\subsection{System Combination and Semi-supervised Learning}
In the last set of experiments, we report on the results of our final system submitted to the ASR challenge organizer. The final system performed an ensemble of the ASR systems previously evaluated in Tables 2 and 3. Specifically, the ASR results of all the abovementioned systems, in the form of word lattices, were first merged (unified) into a single word lattice with equal prior weights. We then conducted Minimum Bayes-Risk (MBR) decoding on the merged lattice, whose outputs were served as the results of our final ASR system. On a separate front, since it was allowed to make use of the label-agnostic evaluation dataset (viz. the corresponding reference transcripts were not provided), we thus went one step further to leverage the label-agnostic evaluation dataset for acoustic model training. That is, we conducted semi-supervised learning of the acoustic model by additionally using the unlabeled evaluation dataset and adopting the strategies proposed in [25] and [26]. As we can see in Table 4, our proposed system-ensemble approach (Row 1) can further improve the best WER results on the development dataset from 18.86\% to 16.70\%. Further, with the additional use of semi-supervised learning, though our best WER result on the development dataset was slightly degraded from 16.70\% to 16.74\%, such combination of the system-ensemble approach with semi-supervised learning achieved a WER result 17.59\% on the evaluation when using our best ASR system configuration. Finally, Table 5 summarizes the final WER results of the participating teams on the evaluation dataset of the TLT-school Challenge.

\begin{table}[]
\centering
\caption{Final WER (\%) results on the evaluation dataset for the participating teams of the TLT-school Challenge.}
\begin{tabular}{cc}
\hline
Participating   Teams                                                                   & WER (\%) \\ \hline
\begin{tabular}[c]{@{}c@{}}ALTA Institute,\\ Cambridge University\end{tabular}          & 15.67    \\ \hline
\begin{tabular}[c]{@{}c@{}}SMIL Lab, National\\ Taiwan Normal\\ University\end{tabular} & 17.59    \\ \hline
Aalto University                                                                        & 18.71    \\ \hline
\begin{tabular}[c]{@{}c@{}}University of\\ Birmingham\end{tabular}                      & 18.80    \\ \hline
Anonymous                                                                               & 19.64    \\ \hline
\begin{tabular}[c]{@{}c@{}}Seoul National University\\ SLPLAB\end{tabular}              & 21.63    \\ \hline
Anonymous                                                                               & 22.24    \\ \hline
Johns Hopkins University                                                                & 26.38    \\ \hline
\begin{tabular}[c]{@{}c@{}}Indian Institute of \\ \\ Technology Bombay\end{tabular}     & 26.61    \\ \hline
Baseline (Organizer)                                                                    & 35.09    \\ \hline
\end{tabular}
\end{table}

\section{Conclusion}
In this paper, we have presented and evaluated the NTNU ASR system participating in the TLT-school Challenge. The promising effectiveness of the joint use of data cleansing, pronunciation modeling, data augmentation, system combination and semi-supervised learning methods for non-native Children’s English speech ASR have been confirmed, through an extensive set of experimental evaluations. As to future work, we plan to apply and extend the aforementioned methods to more sophisticated DNN-HMM or end-to-end ASR systems, as well as other resource-poor ASR tasks.



\end{document}